# Queue Theory based Response Time Analyses for Geo-Information Processing Chain

Jie Chen, Jian Peng, Min Deng, Chao Tao, Haifeng Li

[a] School of Civil Engineering, School of Geosciences and Info-Physics, Central South University, 22

Shaoshan South Road, Changsha, 410076, P.R.China

Abstract: Typical characteristics of remote sensing applications are concurrent tasks, such as those found in disaster rapid response. The existing composition approach to geographical information processing service chain, searches for an optimisation solution and is what can be deemed a "selfish" way. This way leads to problems of conflict amongst concurrent tasks and decreases the performance of all service chains. In this study, a non-cooperative game-based mathematical model to analyse the competitive relationships between tasks, is proposed. A best response function is used, to assure each task maintains utility optimisation by considering composition strategies of other tasks and quantifying conflicts between tasks. Based on this, an iterative algorithm that converges to Nash equilibrium is presented, the aim being to provide good convergence and maximise the utilisation of all tasks under concurrent task conditions. Theoretical analyses and experiments showed that the newly proposed method, when compared to existing service composition methods, has better practical utility in all tasks.

**Keywords**: Geo-Information services chain; services composition; response time; queue theory; concurrencies.

# 1 Introduction

Web services "have a transformative effect on scientific communities, making tools which used to be merely accessible to the specialist available to all, and permitting previous manual data processing and analysis tasks to be automated" {Foster, 2008 #11215}. Web services optimization composition in terms of Quality of Service (QoS) properties, however, is a key problem{Menascé, 2002 #340}. Concurrency of a large number of tasks often exists in Web services-based applications, especially in crisis-orientated management. However, when every task "selfishly" seeks for the optimum solution without considering the performance of the entire service system, such methods will result in conflicts between tasks because numerous concurrent tasks will compete for limited optimal resources.

This means that many tasks will be assigned to the same optimal service at the same time, which results in the degeneration of processing service capability, and cause service quality decline in all service chains{Alameh, 2003 #112;Alameh, **2002** #111}. Each service must deal with different tasks under concurrency; thus, these tasks form a waiting queue, and response time is not only influenced by the process ability of the process service itself, but also by the task load. Moreover, the complicated construction of service chain control flow makes the calculation of QoS aggregation value of service chain, particularly in terms of response time, much harder{Zeng, 2003 #1231;Zeng, 2004 #475;Ko, 2008 #9360}.

The response time of a service chain contains two factors: expected value and variance. The variance represents the stability of the response time of a service, which is an important consideration in the service selection.

Unfortunately, all existing response time calculate methods for service chain only consider single task {Menascé, 2002 #340;Bilgin, 2004 #960;Zeng, 2004 #475;Canfora, 2004 #965;Canfora, 2005 #1222;Claro, 2005 #194;Canfora, 2008 #10582}. None of them take the queue waiting time cause by concurrent tasks into account. This leading to the impreciseness of response time computing.

Hence, we proposal a queue theory based response time computing method for web services chain. We first analyze the control flow and task arrive model in Web services chain, then model the waiting process in the Web services concurrent task as a queue model. Final, we describe queue theory based response time computing method for each control flow style.

Further content organization is represented below: the 2nd section discusses the basic control flow model and query model for Web services chain; the 3rd section describes queue theory based response time computing method for web services chain; the 4th summarizes all the work in the paper, and forecasts the further work.

# 2 Basic model

Control flow model. OWL-S{Ankolekar, 2005 #134;Li, 2008 #10536} define control flow as Sequence, Split, Condition, Split+Join, Unordered, Choice, Iterate, If-Then-Else, Repeat-Until, and Repeat-While. But it's redundancy in the definition, and can be included as following:

Sequence Relation: giving services $w_1$ and $w_2$, if the execution of $w_2$ is after $w_1$, then $w_1$ and $w_2$ are sequence relation.

Parallel Relation: giving services $w_1$ and $w_2$, if the executions of $w_1$ and $w_2$ are independent of each other, then $w_1$ and $w_2$ are parallel relation.

Fork Relation: giving services $w_1$ and $w_2$, if either executing $w_1$ or $w_2$ according to execution result, then $w_1$ and $w_2$ are fork relation.

Iteration Relation: giving service $w_1$, if $w_1$ is executed continuously, then $w_1$ is iteration relation.

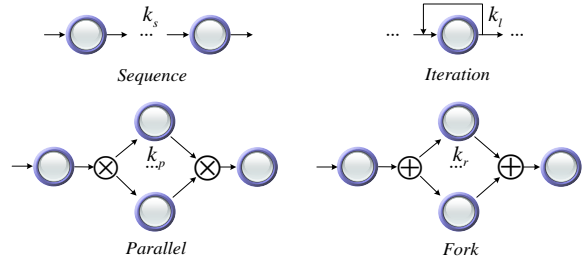

Figure 1. Control flow model

**Queue model.** Under multitask concurrency, tasks may have to wait in the service execution queue. Thus, the computing method of the response time is also different. The queue model of $M/M/1$ (Figure 2) is applied to estimate the response time of processing services. The cost and availability can be regarded as uninfluenced by the continual concurrent tasks; accordingly, we can utilize the same aggregation algorithm as in single task conditions{Zeng, 2004 #475} (Table I).

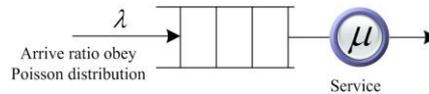

Figure 2. $M/M/1$ queue model.

TABLE I. RESPONSE TIME DIMENSION AGGREGATION METHODS

| Control structure | Response time |
|---|---|
| Sequential structure | $\sum_{l=1}^{L} 1/(\mu^l - \lambda)$ |
| Parallel structure | $\max_{m} \sum_{l=1}^{L} 1/(\mu_m^l - \lambda)$ |
| Iteration structure | $\sum_{l=1}^{L} 1/(p\mu^l - \lambda)$ |
| Branch structure | $\sum_{n=1}^{N} \sum_{l=1}^{L} 1/(\mu_n^l - b_n\lambda)$ |

# 3 Queue theory based response time computing method

## 3.1 Atmotic structure

In the $M/M/1$ queue model, every task $i$'s space of arrival time follows the exponential distribution with the speed of $\lambda_i$, and every processed service time follows the exponential distribution with the parameter of $\mu$; therefore, every processing service time is{Gross, 1985 #11195}:

$$W = 1/(\mu - \lambda_i) \quad (1)$$

To calculate the aggregation response time of the service chain, analysis should be carried out in sequential, parallel, branch, and iteration structures. Sequential structure represents the $L$ abstract services implemented by order; the iteration structure shows that the abstract services within will be re-implemented for $K$ times; the parallel structure states that the $M$ branched abstract services must be simultaneously carried; and the branch structure means there are $N$ branches, among which every branch $n$ is selected to be implemented according to the possibility $b_n$.

Sequential structure. In the queue network, according to Burke's theorem{Bacon, 1994 #1218}, for the $M/M/1$ queue with the arriving rate of $\lambda$, the output is also a Poisson process with the rate of $\lambda$. That is to say, for all the processing services of the sequential structure, the arrival and departure processes follow the Poisson distribution. As a result, the computing method of the response time of the sequential structure is:

$$W = \sum_{l=1}^{L} 1/(\mu^l - \lambda) \tag{2}$$

Where $L$ represents the total length of the sequential structure, and $l$ indicates the index of steps in the sequential structure (Figure 3).

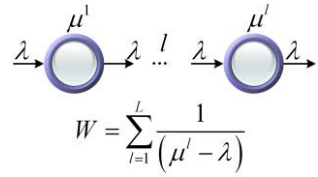

Figure 3. Response time computing model of sequential structure.

Parallel structure. In the parallel structure, every branch needs to be implemented; therefore, the arrival rate of every branch task is $\lambda$, based on Burke's theorem{Bacon, 1994 #1218}. The departure process of every branch also follows the Poisson distribution with the rate of $\lambda$ (Figure 4). Furthermore, in parallel structure, the total response time is determined by the longest parallel branch, i.e., the key path{Zhu, 2009 #11677}. As a result, to solve the response time of the parallel structure, the key path should first be solved, then the parallel structures serialized (Figure 4). Hence, the computing method of the response time of the parallel structure is:

$$W = \max_{M} \sum_{l=1}^{L} 1/(\mu_m^l - \lambda) \tag{3}$$

Where $k_p$ represents the number of parallel branches, and $m$ indicates the index of parallel structure branches.

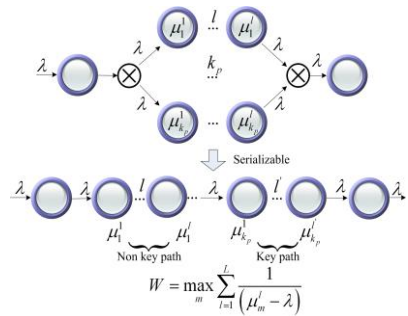

Figure 4. Response time computing model of parallel structure.

(3) Branch structure. Branch structure describes the possibility of execution route being selected, if there are $N$ branches, and every branch $n$'s possibility of being chosen is $b_n$ and satisfies $\sum_{n=1}^{N} b_n = 1$. Accordingly, the arrival rate of every branch task is $b_n \lambda$. Based on Burke's theorem, the departure of every branch follows the Poisson distribution with the rate of $b_n \lambda$ (Figure 5). Tasks are allocated to different branches with different possibilities in the branch structure; thus, we can still use the serialization method to calculate the response time (Figure 5). The response time of the parallel structure is computed as:

$$W = \sum_{n=1}^{N} \sum_{l=1}^{L} 1/(\mu_n^l - b_n \lambda) \tag{4}$$

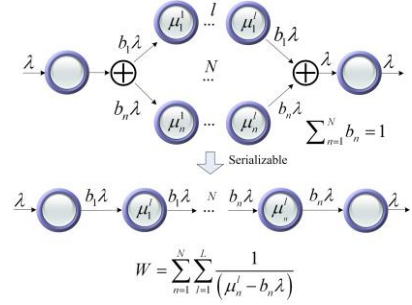

Figure 5. Response time computing model of branch structure.

(4) Iteration structure. Different from the loop peeling[6] and unfolding[10] methods, we consider the iteration structure as the feedback to execution in the queue model (Figure 6). Even if it does not obey Poisson distribution inside the iteration structure, the behaviors of internal processing services can still be independent as $M/M/1$ because of the feedback. As a consequence, the iteration structure is still a part of the $M/M/1$ queue network. In accordance with Jackson theorem, presuming the internal arrival rate of the iteration structure is $r$, the feedback possibility is $1-p$, consequently:

$$r = \lambda + (1-p)r, \quad r = \lambda/p \tag{5}$$

And the response time in the iteration structure is:

$$W = \sum_{l=1}^{L} 1/\mu_l - \lambda/p = \sum_{l=1}^{L} p/(p\mu_l - \lambda) \tag{6}$$

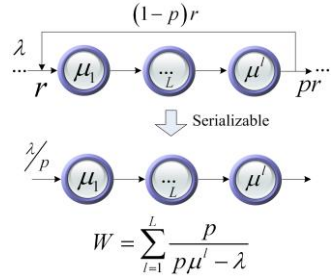

Figure 6. Response time computing model of iteration structure.

Under the circumstances of multitask concurrency, every processing service needs to simultaneously deal with different tasks; thus, for every processing, the arrival quantity of actual tasks is $\sum_{m=1}^{I} s_{m,j}^l \lambda_m$, and the entire expected response time of task $i$ is the aggregation formulas of every QoS dimension (TABLE II). When it comes to cost and availability, the computing methods and materials are the same with {Zeng, 2004 #475}.

Sequential structure: $\sum_{l \in S} \sum_{j=1}^{J} \dfrac{s_{i,j}^l}{\left(\mu_j^l - \sum_{m=1}^{I} s_{m,j}^l \lambda_m\right)}$

Parallel structure: $\max_{cp \in P} \sum_{l \in cp} \sum_{j=1}^{J} \dfrac{s_{i,j}^l}{\left(\mu_j^l - \sum_{m=1}^{I} s_{m,j}^l \lambda_m\right)}$

Branch structure: $\sum_{k_n=1}^{k_r} \sum_{l \in C_{k_n}} \sum_{j=1}^{J} \frac{s_{i,j}^l}{\left(\mu_j^l - b_{k_n} \sum_{m=1}^{I} s_{m,j}^l \lambda_m\right)}$

Iteration structure: $\sum_{l \in C} \sum_{j=1}^{J} \frac{s_{i,j}^l}{\left(p\mu_j^l - \sum_{m=1}^{I} s_{m,j}^l \lambda_m\right)}$

Note: in the response time function, $\eta = 1/p$ represents the feedback possibility; in the cost and availability function, $\eta = \bar{p}$ represents expected recycle times. $cp$ represents the key path.

TABLE II. RESPONSE TIME AGGREGATE FUNCTION

| Control structure | Response time($T_i$) |
| --- | --- |
| Sequential($S$) | $\sum_{l \in S} \sum_{j=1}^{J} \frac{s_{i,j}^l}{(\mu_j^l - \sum_{m=1}^{I} s_{m,j}^l \lambda_m)}$ |
| Parallel($P$) | $\max_{cp \in P} \sum_{l \in cp} \sum_{j=1}^{J} \frac{s_{i,j}^l}{(\mu_j^l - \sum_{m=1}^{I} s_{m,j}^l \lambda_m)}$ |
| Iteration ($C$) | $\sum_{l \in C} \sum_{j=1}^{J} \frac{s_{i,j}^l}{(\mu_j^l - \eta \sum_{m=1}^{I} s_{m,j}^l \lambda_m)}$ |
| Branch($R$) | $\sum_{k_n=1}^{k_r} \sum_{l \in C_{k_n}} \sum_{j=1}^{J} \frac{s_{i,j}^l}{(\mu_j^l - b_{k_n} \sum_{m=1}^{I} s_{m,j}^l \lambda_m)}$ |

Theorem 1. Parallel, branch, and iteration structures in the service chain can be serialized with equal values to attain a sequential structure without changing the values of aggregate function.

Proof: Although cost aggregate function is mainly for summation calculation, and the availability function is for multiplication, these can be converted into summation calculations via logarithmic function. For the response time (excluding parallel structure), it is the maximum or minimum value calculation. However, after finding the key path, the maximum or minimum value calculation can still be converted to summation computation. Therefore, these calculations are linear structures and can be serialized.

The significance of the serialization lies in its removal of the different expressions of QoS indicator aggregate functions due to the difference in control structure, and in building a unified model. Presuming the service chain after serialization is $L$, the general characteristics are not lost; let the length be $L$ (as the length of the service chain after being serialized), and the structure factor $\kappa$ introduced to unify the expression.

Definition 1: Structure factor $\kappa$ depicts the equalization value of the serialization of different control structures.

$$\kappa = \begin{cases} 1, N = 1 & if\ l \in S\ or\ l \in P \\ \eta, N = 1 & if\ l \in C \\ b_{k_n} & if\ l \in R \end{cases} \quad (7)$$

$$T_i(\mathcal{L}) = \sum_{l \in cp}^{L} \sum_{j=1}^{J} \frac{s_{i,j}^l}{\left(\mu_j^l - \kappa \sum_{m=1}^{I} s_{m,j}^l \lambda_m\right)} \quad (8)$$

# 4 Conclusion

A queue theory based response time computing method for web services chain is presented. It's a new way to calculate the response time for Web services chain, especially for concurrent tasks situation. In this paper, we model each services as M/M/1 queue, and we will introduce more complex queue model for more precision in the future work.